\documentclass[pre,twocolumn,showpacs,preprintnumbers,floatfix]{revtex4}
\usepackage{amsfonts}
\usepackage{makeidx}
\usepackage{amssymb}
\usepackage{graphicx}

\begin{document}

\title{\textbf{Quasiperiodic graphs at the onset of chaos}}
\author{B. Luque$^{1}$, M. Cordero-Gracia$^{1}$, M. G\'{o}mez$^{1}$, and A. Robledo$%
^{2}$}
\affiliation{$^1$ Dept. Matem\'{a}tica Aplicada y Estad\'{\i}stica. ETSI Aeron\'{a}%
uticos, Universidad Polit\'{e}cnica de Madrid, Spain.\\
$^{2}$ Instituto de F\'{\i}sica y Centro de Ciencias de la Complejidad,
Universidad Nacional Aut\'{o}noma de M\'{e}xico, Mexico.}

\begin{abstract}
We examine the connectivity fluctuations across networks obtained when the
horizontal visibility (HV) algorithm is used on trajectories generated by
nonlinear circle maps at the quasiperiodic transition to chaos. The
resultant HV graph is highly anomalous as the degrees fluctuate at all
scales with amplitude that increases with the size of the network. We
determine families of Pesin-like identities between entropy growth rates and
generalized graph-theoretical Lyapunov exponents. An irrational winding
number with pure periodic continued fraction characterizes each family. We
illustrate our results for the so-called golden, silver and bronze numbers.
\end{abstract}

\maketitle

%\date{ }

\section{Introduction}

The onset of chaos is a prime dynamical phenomenon that has attracted
continued attention motivated by the aim to both expand its understanding
and to explore its manifestations in many fields of study \cite{strogatz1}.
From a theoretical viewpoint, chaotic attractors generated by
low-dimensional dissipative maps have ergodic and mixing properties and, not
surprisingly, they can be described by a thermodynamic formalism compatible
with Boltzmann-Gibbs (BG) statistics \cite{dorfman1}. But at the transition
to chaos, the infinite-period accumulation point of periodic attractors,
these two properties are lost and this suggests the possibility of exploring
the limit of validity of the BG structure in a precise but simple enough
setting. The horizontal visibility (HV) algorithm \cite{luque1,luque2}
that transforms time series into networks has offered \cite{luque3,luque4,luque5,luque6,luque7}
a view of chaos and its genesis in low-dimensional maps from
an unusual perspective favorable for the appreciation and understanding of
basic features. Here we present the scaling and entropic properties
associated with the connectivity of HV networks obtained from trajectories
at the quasiperiodic onset of chaos of circle maps \cite{hilborn1} and show
that this is an unusual but effective setting to observe the universal
properties of this phenomenon.

The three well-known routes to chaos in low-dimensional dissipative systems,
period-doubling, intermittency and quasiperiodicity, have been analyzed
recently \cite{luque3,luque4,luque5,luque6,luque7} via the HV formalism, and complete sets
of graphs, that encode the dynamics of all trajectories within the
attractors along these routes, have been determined. These graphs display
structural and entropic properties through which a distinct characterization
of the families of time series spawned by these deterministic systems is
obtained. The quantitative basis for these results is provided by the
corresponding analytical expressions for the degree distributions. The graph
at the transition to chaos has been studied only for the period-doubling
route for which connectivity expansion an entropy growth rates
have been determined and found to be linked by Pesin-like identities \cite{luque5}. Here
we present results for the transition to chaos for the quasiperiodic route
that expand on this findings and suggest that structural and entropic
properties of such networks are linked by Pesin-like equalities that use
generalizations of the ordinary Lyapunov and BG entropy expressions.

We refer to Pesin-like identities as those that were first found to occur at the period-doubling transitions to chaos that link generalized Lyapunov exponents to entropy growth rates at finite, but all, iteration times \cite{baldovin1,mayoral1}. Recently \cite{luque5} these identities were retrieved in a network context via the HV method. Pesin-like identities differ from the genuine Pesin identity, the single positive Lyapunov exponent version of the Pesin theorem \cite{pesin1}, for chaotic attractors in one-dimensional iterated maps. The Pesin identity links asymptotic quantities that are invariant under coordinate transformations, whereas the finite-time Pesin-like identities, that appear for vanishing ordinary Lyapunov exponent are coordinate dependent. However, in the case of period doubling it has been seen that the identities remain valid when different coordinate systems are used to determine them, as in Refs. \cite{luque5} and \cite{baldovin1}.

The rest of this paper is as follows: We first recall the HV algorithm \cite%
{luque1,luque2} that converts a time series into a network and focus
on the quasiperiodic graphs \cite{luque6} as the specific family of HV
graphs generated by the standard circle map. We then expose the universal
scale-invariant structure of the graphs that arise at the infinite period
accumulation points by focusing on the golden ratio route. We describe the
diagonal structure of these graphs when represented by the exponential of
the connectivity, and introduce a generalized graph-theoretical Lyapunov
exponent appropriate for the subexponential growth of connectivity
fluctuations. Subsequently, we show how the collapse of the diagonal
structure into a single one represents the scale-invariant property that
governs the degree fluctuations. Following this, we analyze the network
expression for the entropy rate of growth and find a spectrum of Pesin-like
identities. Finally, we show that all the previous results can be
generalized by considering winding numbers given by any quadratic
irrational. We discuss our results.

\begin{figure*}[t]
\includegraphics[width=1.0\columnwidth,angle=0,scale=1.8]{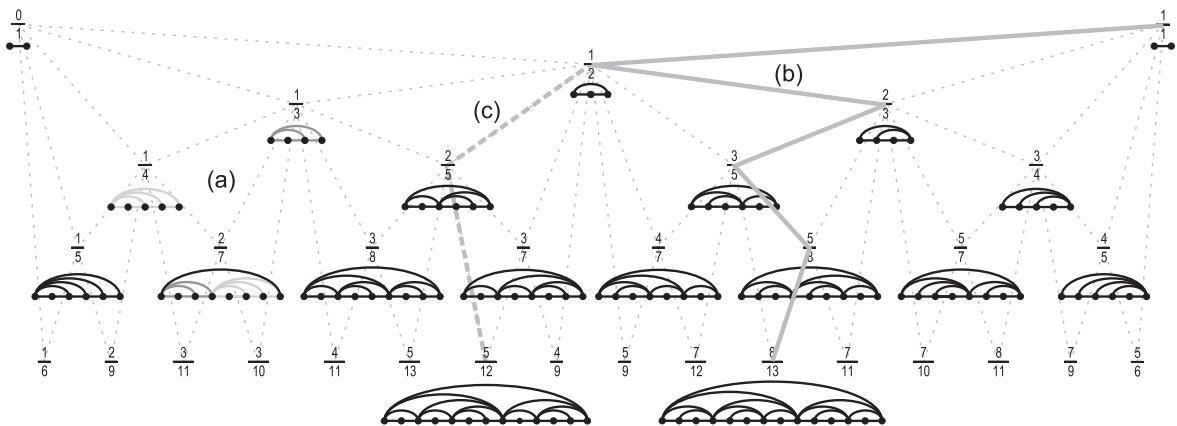}
\caption{{\protect\small Six levels of the Farey tree and the periodic
motifs of the graphs associated with the corresponding rational fractions $%
p/q$ taken as dressed winding numbers $\protect\omega $ in the circle map
(for space reasons only two of these are shown at the sixth level). (a) In
order to show how graph inflationary process works, we have highlighted an example
using different grey tones on the left side. See Ref. \protect\cite{luque6}
for details. (b) First five steps along the Golden ratio route, $b=1$ (thick solid
line); (c) First three steps along the Silver ratio route, $b=2$ (thick dashed line).}
}
\label{farey}
\end{figure*}

\section{Quasiperiodic graphs at the golden ratio onset of chaos}

The idea of extracting graphs from time series is hardly new and over the past years several approaches have been proposed and are currently developed \cite{Crutchfield, zhang06, kyriakopoulos07, xu08, donner10, donner11, donner11-2, campanharo11}. The HV approach is chosen here because of both its simplicity of implementation and its capability to produce analytical results in closed form for quantities that are generally difficult to determine. As we see below this is corroborated for the present enterprise. For the circle map it has been possible to determine previously the relevant dynamical quantities at the transition to chaos only for the golden route \cite{robledo1}. In contrast, in the present study it has been possible to generalize this result effortlessly for an infinite number of routes to chaos associated with all the quadratic irrational numbers.

%The idea of extracting graphs from time series is hardly new. Already in 1989 Crutchfield and Young \cite{Crutchfield} proposed the construction of discrete finite automata (epsilon machine) from symbolic dynamics series in order to determine the precise form of the complexity of an iterated map. Recently the idea has progressed with renovated strength to develop new approaches \cite{zhang06, kyriakopoulos07, xu08, donner10, donner11, donner11-2, campanharo11}.\\

The horizontal visibility (HV) algorithm is a general method to convert time
series data into a graph \cite{luque1,luque2} and is minimally
stated as follows: assign a node $i$ to each datum $\theta _{i}$ of the time
series $\{\theta _{i}\}_{i=1,2,...}$ of real data, and then connect any
pair of nodes $i$, $j$ if their associated data fulfill the criterion $%
\theta _{i}$ , $\theta _{j}$ $>$ $\theta _{n}$ for all $n$ such that $i<n<j$.
We note that the HV algorithm is related to the permutation entropy scheme \cite{Bandt} in which the problem of the partition of symbols of a time series is sorted out by simple comparison of nearest-neighbor values within the series. The HV method addresses in a similar way this problem, but in addition it makes use of comparisons of values between neighbors that can be separated by long distances, and consequently it stores additional information of the series in the structure of the resulting HV graph.

The HV method has been applied \cite{luque6} to trajectories generated by
the standard circle map \cite{Landau, Ruelle, Shenker, Kadanoff, Rand, Rand2, hilborn1} given by
\begin{equation}
\theta _{t+1}=f_{\Omega ,K}(\theta _{t})=\theta _{t}+\Omega -\frac{K}{2\pi }%
\sin (2\pi \theta _{t}),\;\textrm{mod}\;1,  \label{circlemap1}
\end{equation}%
representative of the general class of nonlinear circle maps: $\theta
_{t+1}=f_{\Omega ,K}(\theta _{t})=\theta _{t}+\Omega +K\cdot g(\theta
_{t}),\;\textrm{mod}\;1$, where $g(\theta )$ is a periodic function that
fulfills $g(\theta +1)=g(\theta )$. This family of maps exhibit universal
properties that are preserved by the HV algorithm \cite{luque6} so that
without loss of generality we explain below our findings in terms of the
standard circle map, where $\theta _{t}$, $0\leq \theta _{t}<1$, is the
dynamical variable, the control parameter $\Omega $ is called \emph{bare
winding number}, and $K$ is a measure of the strength of the nonlinearity.
The \emph{dressed winding number} for the map is defined as the limit of the
ratio: $\omega \equiv \lim_{t\rightarrow \infty }(\theta _{t}-\theta _{0})/t$%
. For $K\leq 1$ trajectories are periodic (locked motion) when the
corresponding dressed winding number $\omega (\Omega )$ is a rational number
$p/q$ and quasiperiodic when it is irrational. For $K=1$ ($\emph{critical}$
circle map) locked motion covers the entire interval of $\Omega $ leaving
only a multifractal subset of $\Omega $ unlocked.\\

The periodic time series of period $q$ that constitutes the trajectory within an attractor
with $\omega (\Omega )=p/q$ is represented in the HV graph by the repeated
concatenation of a motif, a number of which are shown in Fig. \ref{farey}. The display
of these motifs in the Farey tree in Fig. \ref{farey} helps visualize the inflationary process
that takes place when the HV network grows at the onset of chaos \cite{luque6}.
For illustrative purposes in Fig. \ref{farey} we show the periodic motifs of
the HV graphs that are associated with the irreducible rational numbers $%
p/q\in \lbrack 0,1]$, and we place them on the Farey tree \cite{hilborn1}
along which routes to chaos take place. A well-studied case
is the sequence of rational approximations of $\omega _{\infty }=\phi
^{-1}=(\surd 5-1)/2=0.618034...$ , the reciprocal of the golden ratio, which
yields winding numbers $\{\omega _{n}=F_{n-1}/F_{n}\}_{n=1,2,3...}$ where $%
F_{n}$ is the Fibonacci number generated by the recurrence $%
F_{n}=F_{n-1}+F_{n-2}$ with $F_{0}=1$ and $F_{1}=1$. The first few steps of
this route can be seen in Fig. \ref{farey}(b).

The trajectories generated by the map with initial condition $\theta _{0}=1$
at the golden ratio onset of chaos define a multifractal attractor that
forms a striped pattern of positions when plotted in logarithmic scales,
i.e. $\ln \theta _{t}$ vs $\ln t$. See Fig. 3 in Ref. \cite{robledo1}.
This attractor corresponds to the accumulation point $\Omega _{\infty
}=\lim_{n\rightarrow \infty }\Omega _{n}$ of bare
winding numbers $\Omega _{n}$ that characterize superstable
trajectories of periods $F_{n}$, $n=1,2,3,...$, $\Omega _{\infty }
=0.606661...$  \cite{robledo1}. A sample of this time series is shown in
the top panel of Fig. \ref{lyapunovcirclemap-2}. In the bottom panel of the
same figure we plot, in logarithmic scales, the outcome of the HV method
with use of the variable $\exp k(N)$, where $k(N)$ is the degree of node $N$
in the graph generated by the time series $\theta _{t}$ (that is, $N\equiv
t=1,2,3,...$). Notice that the distinctive striped pattern of the attractor \cite{robledo1}
is present in the figure, although in a simplified manner where the fine
structure is replaced by single lines of constant degree. The HV algorithm
transforms the multifractal attractor into a discrete set of connectivities.

\section{Diagonal structure of the connectivity fluctuations}

\begin{figure*}[t]
\includegraphics[width=1.0\columnwidth,angle=0,scale=1.8]{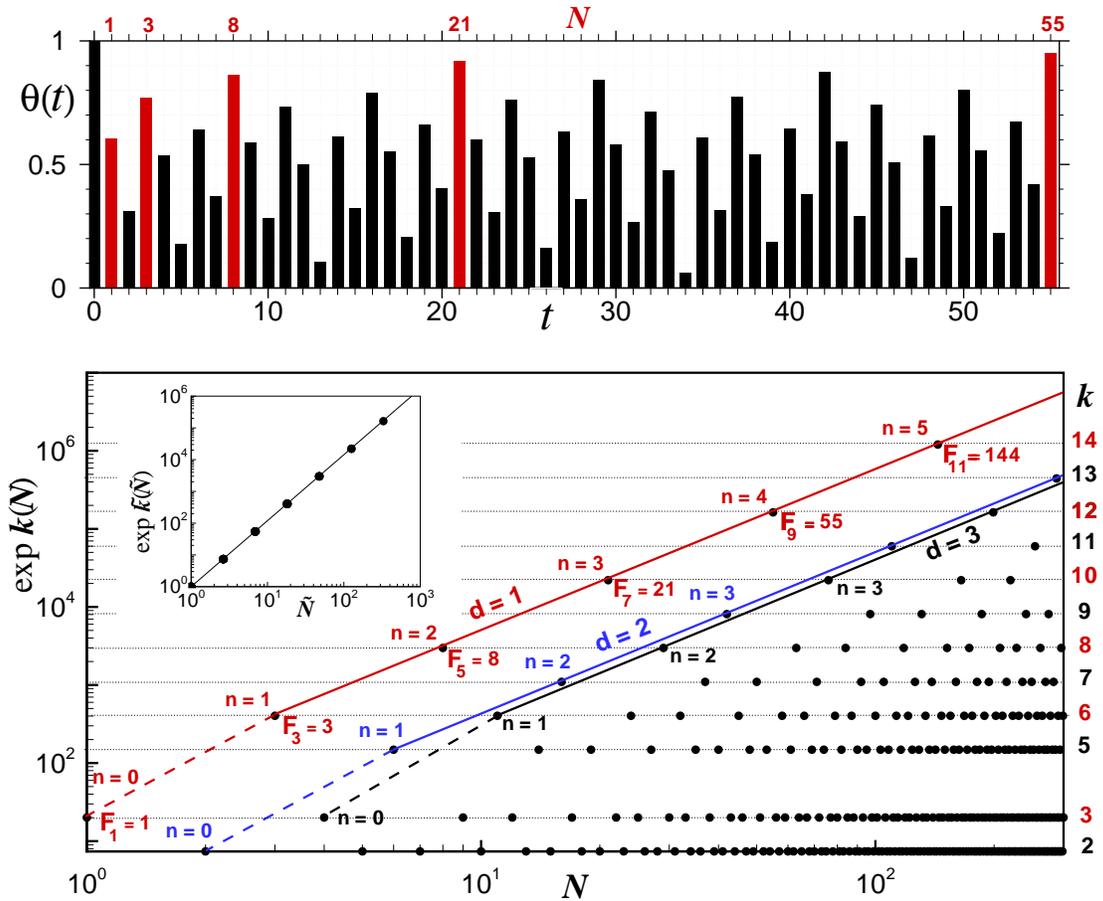}
\caption{ {\protect\small Top: Positions }$\protect\theta _{t}$
{\protect\small as a function of }${\protect\small t}${\protect\small \ for
the first }${\protect\small 55}${\protect\small \ data for the orbit with
initial condition }${\protect\small \protect\theta }_{0}{\protect\small =1}$%
{\protect\small \ at the golden ratio onset of chaos\ (see text) of the critical
circle map }${\protect\small K=1}${\protect\small . The data highlighted are
associated with specific subsequences of nodes (see text). Bottom: Log-log
plot of }${\protect\small \exp k(N)}${\protect\small \ as a function of the
node $N$ for the HV graph generated from same time series as as for the upper
panel but for }$3\times10^{2}${\protect\small \ iterations,
where }${\protect\small N=t}${\protect\small . The distinctive band pattern
of the attractor manifests through a pattern of single lines of constant
degree. The node positions of some node subsequences along diagonals is
highlighted as guide lines to the eye. The inset shows the collapse of all
nodes in the graph into a single diagonal (see text). }}
\label{lyapunovcirclemap-2}
\end{figure*}

It is clear from the bottom panel of Fig. \ref{lyapunovcirclemap-2} that the
degree $k(N)$, and also $\exp k(N)$, fluctuates when $N$\ is increased step
by step via a deterministic pattern of ever increasing amplitude. Notice
also in the same panel the diagonal lines that are drawn to connect
sequences of node-connectivity ($N,k$) values; there is a main diagonal
followed by two other diagonals close to each other. These ($N,k$)
sequences fall asymptotically along parallel straight lines, that begin after the initial
steps from the lowest values of the degree, $k=2$ or $k=3$, skip the absent $k=4$,
and reach the values $k=5$ or $k=6$, and therefore the sequences obey a power law with the
same exponent. There are many more sequences along same-slope diagonals, not
highlighted in the figure, arranged in close groups and that trace all other
possible connectivities $k(N)$. See also Fig. 3 in Ref. \cite{robledo1}. It
is by examining the dependence of $k(N)$ along each member of this family of
diagonals that the scaling and entropic properties of the network are
determined.\\

\begin{figure}[t]
\includegraphics[viewport= 0 150 800 550,clip, width=\columnwidth,angle=0,scale=1]{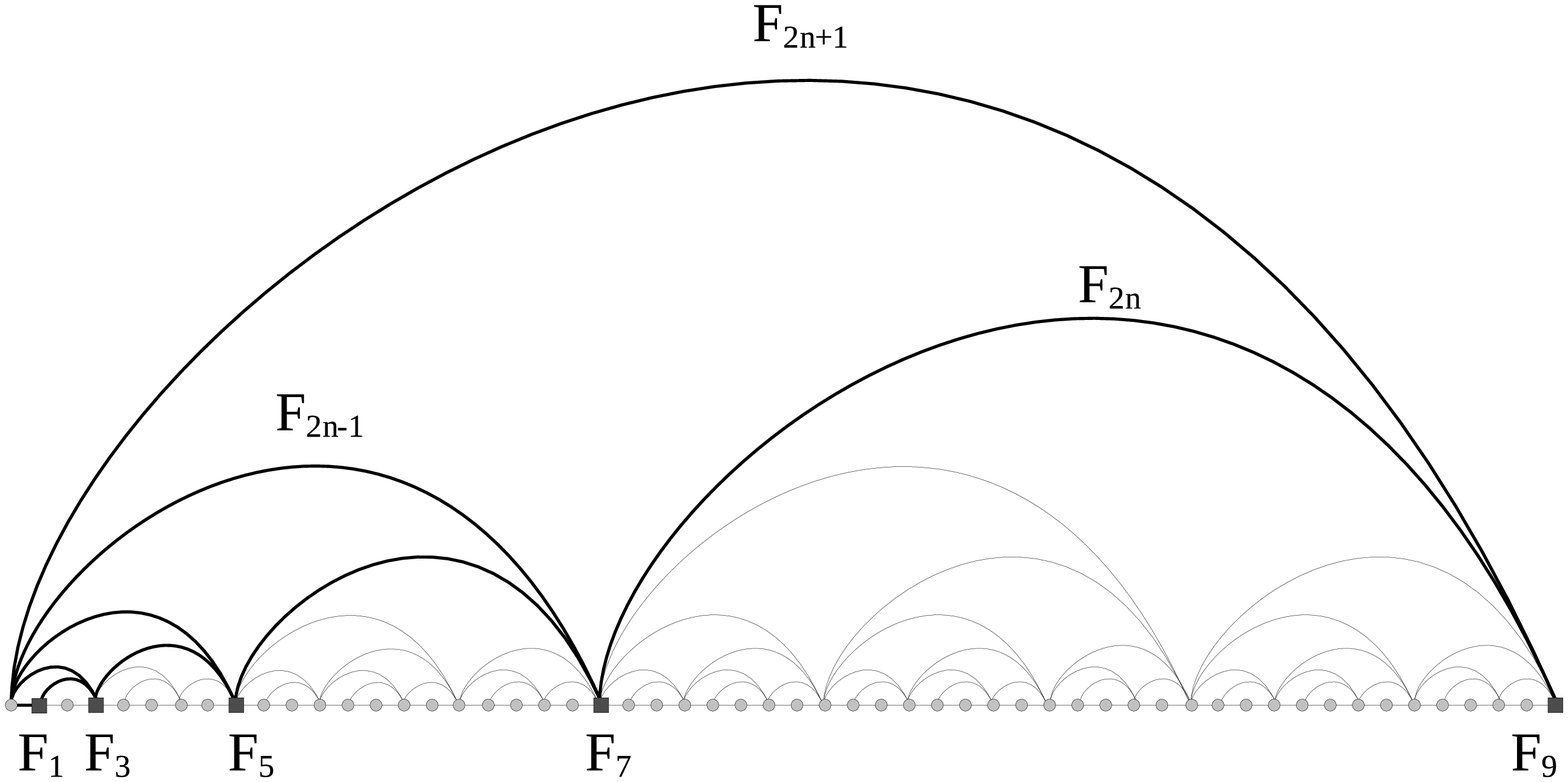}
\caption{ {\protect\small {First substructures of the quasiperiodic graph
associated with the golden ratio route to chaos. The resulting patterns follow
from the universal order with which an orbit visits the positions of the
attractor. The quasiperiodic graph associated with the time series generated
at the onset of chaos ($n\rightarrow \infty $) is the result of an infinite
application of the inflationary process by which a graph at period $F_{2n+1}$
is generated out of graphs at periods $F_{2n}$ and $F_{2n-1}$ \protect\cite%
{luque6}. The first few node/time steps along the first diagonal ($d=1$) are
highlighted.}}}
\label{golden_graph}
\end{figure}

Thus, the ($N,k$) pairs in the graph define a structure in diagonals $%
d=1,2,3,...$, and on each diagonal $d$ we label the particular nodes that
lie on it as $n=0,1,2,...$ Thus, $N(n;d)$ indicates the node/time for the $n$-th position on diagonal $d$. For example, in the first and main diagonal $d=1$ in Fig. 2 we have $N(0;1)=1=F_1$, $N(1;1)=3=F_3$, $N(2;1)=8=F_5$, $N(3;1)=21=F_7$,...
As it can be seen in the top panel of Fig. \ref{lyapunovcirclemap-2}, the matching positions
$\theta_{t}$, $t=F_{2n+1}$ (highlighted) grow monotonically when removed from the rest of the time series,
and according to the HV algorithm this implies increasing values for the degrees of their corresponding nodes.
For $d=2$ (the second diagonal in Fig. 2) $N(0;2)=2$, $N(1;2)=6$, $N(2;2)=16$, $%
N(3;2)=42 $,... All the nodes $N(n;d)$ can be expressed via the recurrence
formula

\begin{eqnarray}
N(0;d) &=&\textrm{mex}\{N(n;i):1\leq i<d,n\geq 0\},  \nonumber \\
N(1;d) &=&2N(0;d)+d,  \nonumber \\
N(n;d) &=&3N(n-1;d)-N(n-2;d),  \label{recurrencia oro}
\end{eqnarray}%
with $d=1,2,...$ and $n=0,1,2,..$., where the term mex stands for
MinimumEXclude value \cite{conway1} that in this case it means the smallest
value of $N$ that has not appeared in the previous diagonals. In \cite%
{fraenkel1} it is demonstrated that every integer $N$ appears only once
under the above recurrence and this \textit{exotic} enumeration occurs in a
natural way \ in the golden ratio route. In fact, all the time labels $n$
along the diagonals $d=1,2,...$ can be expressed as Fibonacci numbers $%
F_{n}^{(d)}=F_{n-1}^{(d)}+F_{n-2}^{(d)}$ with different initial conditions
for each one of them,%
\begin{eqnarray}
F_{0}^{(d)} &=&d,\ F_{1}^{(d)}=N(0;d),  \nonumber \\
N(n;d) &=&F_{2n+1}^{(d)}.  \label{Fibonacci oro}
\end{eqnarray}%
This recurrence is the consequence of the inflationary process that takes
place in the generation of graphs via the golden ratio route \cite{luque6}. Notice that
this route goes through successive approximants of the continued fraction $%
[1,1,1,...]$ (see Fig. \ref{farey}b). These approximants permanently alternate from larger to
smaller to larger values around the golden number, such that an approximant
graph is generated by concatenation of the two preceding approximant graphs
alternating the order of concatenation at each stage. This can be seen
explicitly in Fig. \ref{golden_graph}.

The recurrence formula in Eq. (\ref{recurrencia oro}) can be solved leading
to an explicit expression convenient for our purposes. First, it can be
demonstrated \cite{fraenkel1} that%
\begin{eqnarray}
N(0;d) &=&\lfloor (d-1)\phi \rfloor +1,  \nonumber \\
N(n;d) &=&\lfloor N(n-1;d)\phi ^{2}\rfloor +1.  \label{oro}
\end{eqnarray}%
Then, use of the approximation $N(n;d)\approx N(n-1;d)\phi ^{2}$ and of the
definition $C_{d}\equiv N(1;d)=\big \lfloor(\lfloor (d-1)\phi \rfloor
+1)\phi ^{2}\big
\rfloor+1$ yields the solution

\begin{equation}
N(n;d)=C_{d}\phi ^{2n-2},\ n\geq 1.  \label{N}
\end{equation}%
This equation captures the values $N(n;d)$ along the diagonals starting
always from $n=1$, that, as we can observe in the bottom panel of Fig. \ref{lyapunovcirclemap-2}%
, are the nodes with connectivities $k=5$ or $k=6$. Furthermore, all the
(parallel straight- line) diagonals can be collapsed into a single one by
first redefining the connectivities in each of them such that the degree is
zero in the initial position $n=1$. To do this it is only necessary to
subtract $5$ or $6$ according to the given diagonal, with the outcome that $%
\widetilde{k}=2n-2$ with $n=1,2,...$ To get the collapse it is sufficient to
introduce the change of variable $\widetilde{N}(n;d)=N(n;d)/C_{d}$ so that $%
\widetilde{N}(n;d)=\phi ^{2n-2}$. We can see the result in the inset in the
bottom panel of Fig. \ref{lyapunovcirclemap-2}. To keep notation simple we
make use of this variable and write $k$ instead of $\widetilde{k}$ from now
on.

\begin{figure}[t]
\centering
\includegraphics[width=\columnwidth]{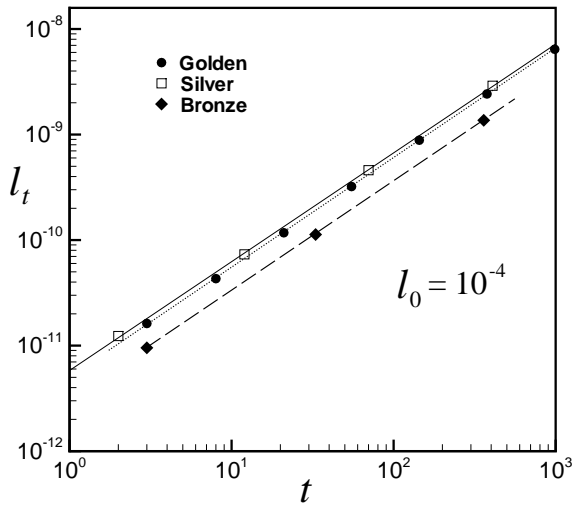}
\caption{{\protect\small Log-log plot of the distance between two nearby
trajectories }$l_{t}=|\protect\theta _{t}-\protect\theta _{t}^{\prime }|$%
{\protect\small \ close to }${\protect\small \protect\theta }_{0}%
{\protect\small =1}${\protect\small , where }${\protect\small l}_{0}%
{\protect\small =10}^{-4}${\protect\small , measured at times }$%
{\protect\small t=N(n;d)}${\protect\small , }${\protect\small n=0,1,2,...}$%
{\protect\small , along the main diagonal }${\protect\small d=1}$%
{\protect\small \ at the transition to chaos for the golden, silver and
bronze routes (see text).}}
\label{expansion}
\end{figure}

\section{Generalized Lyapunov exponents at the accumulation point of the
golden ratio route to chaos}

We define now a connectivity expansion rate for the graph under study. The
formal network analog of the sensitivity to initial conditions in the map is
\cite{luque5}%
\begin{equation}
\xi (N(n;d))\equiv \frac{\exp (k(n))}{\exp (k(1))}=\exp (k(n)),
\label{sensitivity1}
\end{equation}%
since $k(1)=k(N(1;d))=0$. That is, we compare the expansion $\exp (k(n))$
with the minimal $\exp (k(1))=1$ occurring always at nodes at positions $%
N(1;d)$.

From Eq. (\ref{N}) we have

\begin{equation}
k(N(n;d))=2n-2=\ln \left( \frac{N}{C_{d}}\right) ^{\frac{1}{\ln \phi }},
\label{degree1}
\end{equation}%
or%
\begin{equation}
\xi (N(n;d))=\left( \frac{N}{C_{d}}\right) ^{\frac{1}{\ln \phi }}.
\label{sensitivity2}
\end{equation}%
The standard network Lyapunov exponent is defined as%
\begin{equation}
\lambda \equiv \lim_{N\rightarrow \infty }\frac{1}{N}\ln \xi (N),
\label{lyapord1}
\end{equation}%
but since Eq. (\ref{sensitivity2}) indicates that the bounds of the
fluctuations of $\xi (N)$ grow with $N$ slower than $\exp N$ we have $%
\lambda =0$, in agreement to the ordinary Lyapunov exponent at the onset of
chaos.

To get a suitable expansion rate that grows linearly with the size of the
network, we deform the ordinary logarithm in $\ln {\xi (N)}=k(N)$ into $%
\ln _{q}\xi (N)$ by an amount $q>1$ such that $\ln _{q}\xi (N)$ depends
linearly in $N$, where $\ln _{q}x\equiv (x^{1-q}-1)/(1-q)$ and $\ln x$ is
restored in the limit $q\rightarrow 1$ \cite{Koelink, robledo2}. And through this
deformation we define the generalized graph-theoretical Lyapunov exponent as

\begin{equation}
\lambda _{q}\equiv \frac{1}{\Delta N}\ln _{q}\xi (N),  \label{Tsallis}
\end{equation}%
where $\Delta N=N(n;d)-C_{d}$ is the node distance or iteration time
duration between an initial node $N(1;d)$ where $d$ is fixed and $N(n;d)$ is
the final node position. From Eq. (\ref{sensitivity2}) we obtain

\begin{equation}
\lambda _{q}(d)=\frac{1}{N-C_{d}}\frac{\left( \frac{N}{C_{d}}\right) ^{\frac{%
1-q}{\ln \phi }}-1}{1-q}=\frac{1}{C_{d}\ln \phi },  \label{lambdaq1}
\end{equation}%
where the degree of deformation $q$\ is found to be $q=1-\ln \phi $. This
way we have determined a spectrum of generalized Lyapunov exponents $\lambda
_{q}(d)$, one for each diagonal $d=1,2,...$ in Fig.2. The largest value is
for the main diagonal, $\lambda _{q}(1)=(C_{1}\ln \phi )^{-1}$, and the
others gradually decrease as $d\rightarrow \infty $.

\section{$q$-deformed entropy expression and Pesin-like identities}

Having obtained the family of generalized Lyapunov exponents $\lambda
_{q}(d) $ from a suitable expansion rate $\ln _{q}\xi (N)$, we proceed to
analyze the entropic properties of the network. At the transition to chaos
for the golden ratio the HV method creates a single network that represents
many different trajectories. Trajectories initiated at different positions
of the attractor produce networks related to each other by a node
translation equal to the number of iterations needed from one initial
position $\theta _{0}^{(1)}$ to reach the second $\theta _{0}^{(2)}$. The
two positions appear in the trajectory initiated at $\theta _{0}=0$ at times
$t_{1}$ and $t_{2}$, $\theta _{0}^{(1)}=\theta _{t_{1}}$ and $\theta
_{0}^{(2)}=\theta _{t_{2}}$, and the node translation is $\delta
N=t_{2}-t_{1}>0$. This shift property can be visualized in Fig. 2, and is
implicated in the derivation of Eq. (\ref{Tsallis}) for $\lambda _{q}(d)$.
But also, trajectories initiated at positions off the attractor, but
sufficiently close to a position of this set generate the same network, as
the HV method distinguishes differences in trajectory positions only when
they surpass threshold values. There is a basic property of trajectories at
the onset of chaos that combines with the previous remark and that can be
used to describe the rate of entropy growth of the network with its size.
This property is that for a small interval of length $l_{0}$ with $\mathcal{N%
}$ uniformly-distributed initial conditions around, say, $\theta _{0}=0$,
all trajectories behave similarly, remain uniformly-distributed at later
times and follow the concerted pattern shown in Fig. 3 in Ref. \cite%
{robledo1}. Studies of entropy growth
associated with an initial distribution of positions with iteration time $t$
of several chaotic maps \cite{latora1} have established that a linear growth
occurs during an intermediate stage in the evolution of the entropy, after
an initial transient dependent on the initial distribution and before an
asymptotic approach to a constant equilibrium value. In relation to this it
was found, both at the period-doubling \cite{baldovin1, mayoral1} and
at the quasiperiodic golden ratio \cite{robledo1} transitions to chaos, that
(i) there is no initial transient if the initial distribution is uniform and
defined around a small interval of an attractor position, and (ii) the
distribution remains uniform for an extended period of time due to the
subexponential dynamics. In Fig. \ref{expansion} we demonstrate this
property by presenting the time evolution of the distance between to nearby
trajectories, say the endpoints of the interval of length $l_{t}$ containing
the $\mathcal{N}$ uniformly-distributed positions at time $t$, for the
golden ratio transition to chaos, and also for other quasiperiodic
transitions to chaos along other routes discussed below.
But the time evolution of the trajectory distances in Fig. \ref{expansion} can also be that
between any pair of adjacent positions in the initial uniform distribution and
therefore the trajectories' distribution remains uniform after continued iterations.

We denote the above-referred distribution by $\pi (t)=1/W(t)$ where $%
W(0)=l_{0}/\mathcal{N}$ is the number of cells that cover the initial
interval $l_{0}$. As stated, all such trajectories give rise to the same HV
graph, and at iteration times, say, of the form $t=N(n;d)$, $n=1,2,3,\dots $%
, the HV criterion assigns $k=2n-2$ links to the common node $N(n;d)$. The
distribution $\pi $ is defined in the map but we can look at its $n$%
-dependence, $\pi (N(n;d))$, if the scaling properties of the network retain
the scaling property of $\pi $ in the map. We can corroborate this and also
that the entropic properties derived from this distribution are connected to
the network Lyapunov exponents described in the previous section. The
scaling property of the network that yields the collapse of the diagonals in
Fig. \ref{lyapunovcirclemap-2} described above implies that the uniform
distributions $\pi $ for the consecutive node-connectivity pairs ($%
N(n;d),2n-2$) and ($N(n+1;d),2(n+1)-2$) along the same diagonal $d$ scale
with the same factors and this leads us to conclude that the $n$-dependence
for these distributions is
\begin{equation}
\pi (N)=W_{n}^{-1}=\exp (-2n+2).  \label{pi1}
\end{equation}%
But since
\begin{equation}
W_{n}=\exp (2n-2)=\left( \frac{N}{C_{d}}\right) ^{\frac{1}{\ln \phi }},
\label{numberW1}
\end{equation}%
the ordinary entropy associated with $\pi $ grows logarithmically with the
number of nodes $N$, $S_{1}\left[ \pi (N)\right] =\ln W_{j}\sim \ln N$.
However, the $q$-deformed entropy
\begin{equation}
S_{q}\left[ \pi (N)\right] \equiv \ln _{q}W_{n}=\frac{1}{1-q}\left[
W_{n}^{1-q}-1\right] ,  \label{q-entropy1}
\end{equation}%
where the amount of deformation $q$ of the logarithm has the same value as
before, grows linearly with $N$, as $W_{n}$ can be rewritten as
\begin{equation}
W_{n}=\exp _{q}[\lambda _{q}\Delta N],  \label{numberW2}
\end{equation}%
with $q=1-\ln \phi $ and $\lambda _{q}(d)=(C_{d}\ln \phi )^{-1}$.
Therefore, if we define the entropy growth rate
\begin{equation}
h_{q}\left[ \pi (N)\right] \equiv \frac{1}{\Delta N}S_{q}\left[ \pi (N)%
\right]   \label{entropyrate2a}
\end{equation}%
we obtain
\begin{equation}
h_{q}\left[ \pi (N)\right] =\lambda _{q}(d),  \label{entropyrate3}
\end{equation}%
a Pesin-like identity at the onset of chaos (effectively one identity for
each subsequence of node numbers , $n=1,2,3,\dots $, given each by a value
of $d=1,2,3,...$).\newline

\begin{figure*}[t]
\includegraphics[width=1\columnwidth,angle=0,scale=1.8]{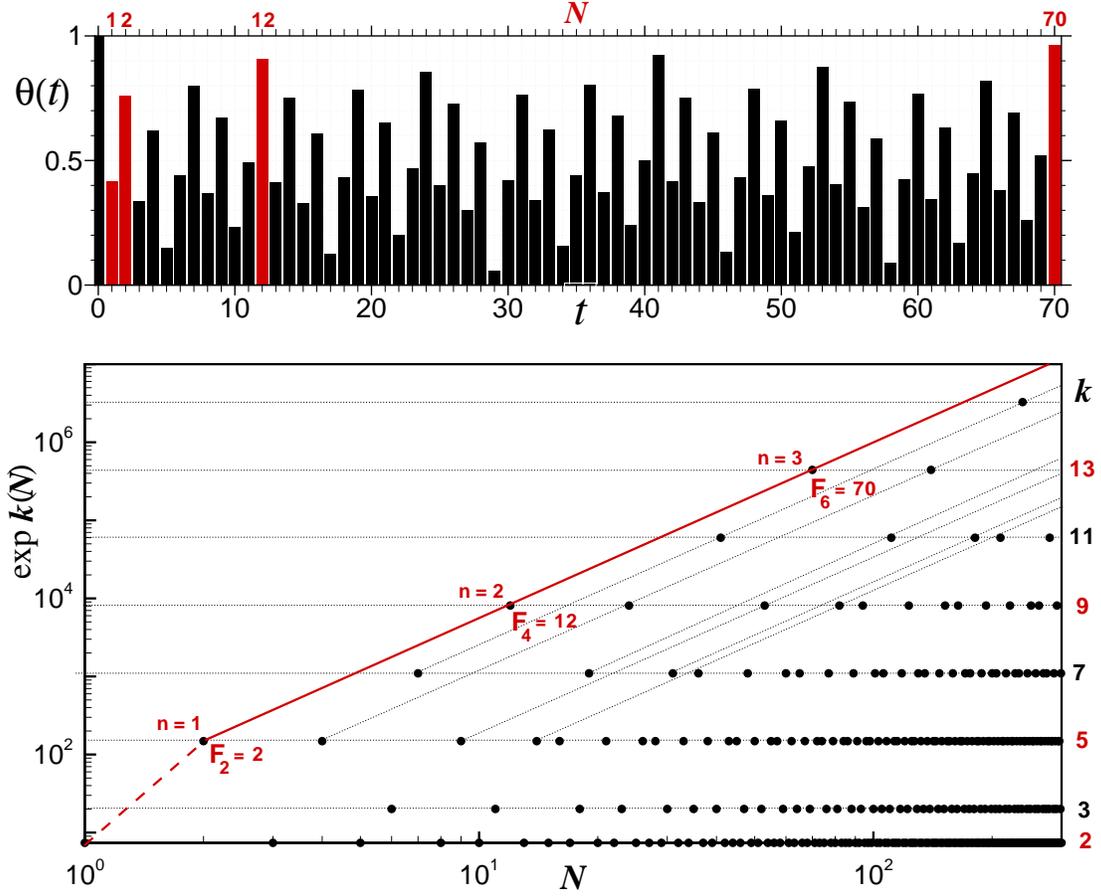}
\caption{ {\protect\small Top: Positions }$\protect\theta _{t}$
{\protect\small as a function of }${\protect\small t}${\protect\small \ for
the first }${\protect\small 70}${\protect\small \ data for the orbit with
initial condition }${\protect\small \protect\theta }_{0}{\protect\small =1}$%
{\protect\small \ at the silver number onset of chaos\ (see text) of the
critical circle map }${\protect\small K=1}${\protect\small . The data
highlighted are associated with specific subsequences of nodes (see text).
Bottom: Log-log plot of }${\protect\small \exp k(N)}${\protect\small \ as a
function of the node }${\protect\small N}${\protect\small \ for the HV graph
generated from same time series as as for the upper panel but for }$3\times10^{2}${\protect\small \ iterations, where }$%
{\protect\small N=t}${\protect\small . The distinctive band pattern of the
attractor manifests through a pattern of single lines of constant degree. The
node positions of some node subsequences along diagonals is highlighted as
guide lines to the eye. }}
\label{plata}
\end{figure*}

\section{Quasiperiodic graphs at the onset of chaos for quadratic irrationals%
}

We can generalize the above results for every quadratic irrational in $[0,1]$
with pure periodic continued fraction representation: $\phi
_{b}^{-1}=[b,b,b,...]=[\overline{b}]$ ($b=1$ , $2$, $3$, correspond to the
golden, silver and bronze routes, respectively). These irrationals are the
solutions of the equation $x^{2}-bx-1=0$, where $b$ is a natural number. The
dressed winding number is now $\omega _{\infty }=\lim_{n\rightarrow \infty
}[1-(F_{n-1}/F_{n})]=\phi _{b}^{-1}$with $F_{n}=bF_{n-1}+F_{n-2}$, $F_{0}=0$,
$F_{1}=1$ and the route to chaos is the infinite sequence of attractors with
periods $F_{n}$, $n=1,2,3,...$(Notice now $F_{n}$ is only a Fibonacci number
when $b=1$). The first few steps of the silver route $b=2$ can be seen in
Fig. \ref{farey}(c), whereas Fig. \ref{plata} shows results for the
attractor at the onset of chaos via this route. Similarly to Fig. \ref%
{lyapunovcirclemap-2} for $b=1$, in the top panel of Fig. \ref{plata} is the
time series for the first $70$ iteration times, while in the bottom panel of
the same figure we plot, in logarithmic scales, the outcome of the HV method
with use of the variable $\exp k(N)$. As it can be observed the networks for
the two cases are qualitatively similar, although there are differences,
mainly the absence of even connectivities when $k>5$.\\

\begin{figure}[t]
\includegraphics[width=\columnwidth,angle=0,scale=1.1]{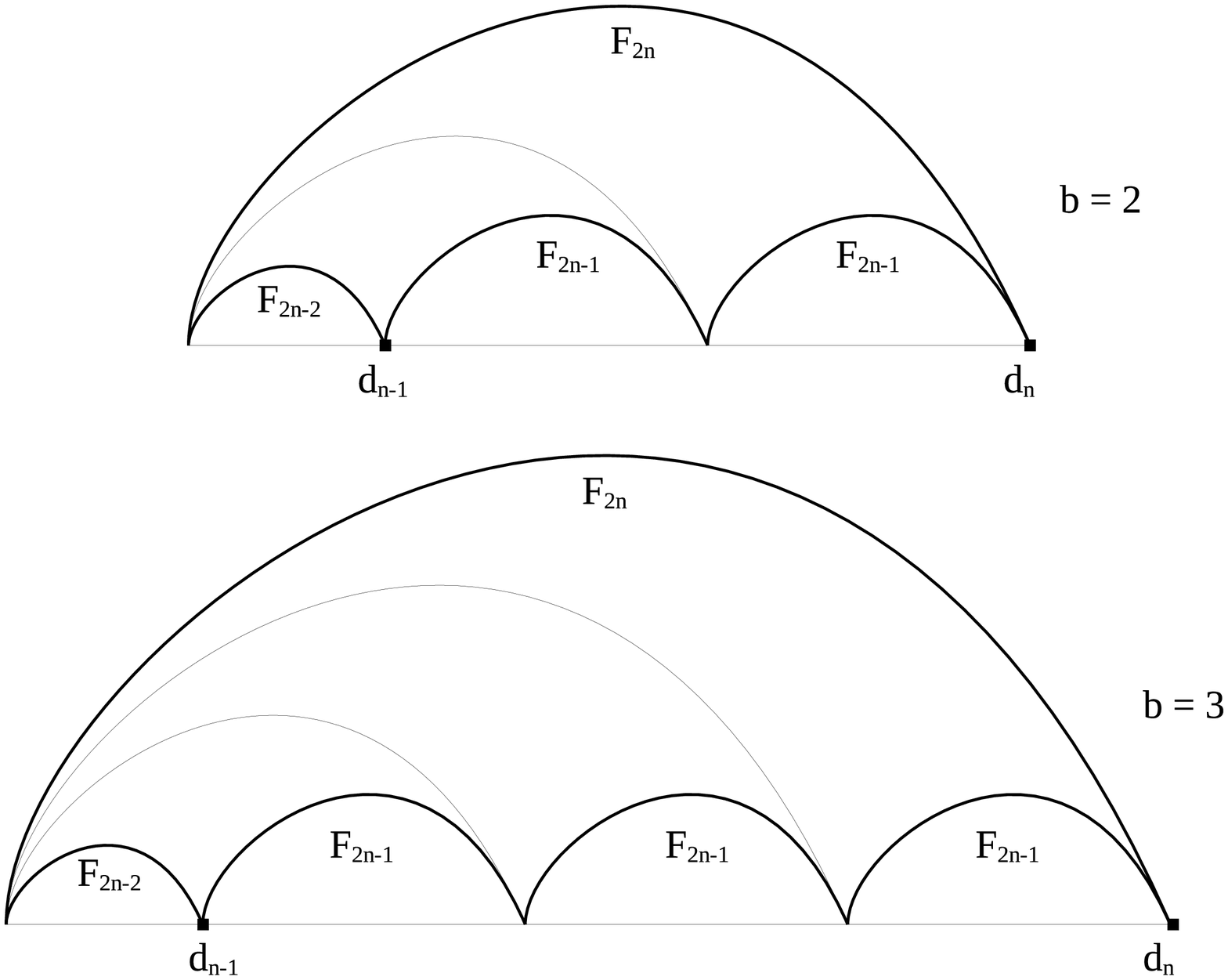}
\caption{ {\protect\small {First substructures of the quasiperiodic graph
associated to (a) the silver number }}${\protect\small b=2}$ {\protect\small %
and (b) the bronze number }${\protect\small b=3}${\protect\small \ {routes
to chaos. The resulting patterns follow from the universal order with which
an orbit visits the positions of the attractors. The quasiperiodic graph
associated with the time series generated at the onset of chaos ($%
n\rightarrow \infty $) is the result of an infinite application of the
inflationary process by which a graph at period $F_{2n}$ is generated out of
graphs at periods $F_{2n-2}$ and $F_{2n-1}$ \protect\cite{luque6}.}} }
\label{metallicinflation}
\end{figure}

This absence can be verified by inspection of the degree distribution $%
P_{\infty }(k)$ for the graphs at the $\omega _{\infty }=\phi _{b}^{-1}$
accumulation points \cite{luque6} \newline

\begin{equation}
P_{\infty }(k)=\left\{
\begin{array}{ll}
\phi _{b}^{-1} & k=2 \\
1-2\phi _{b}^{-1} & k=3 \\
(1-\phi _{b}^{-1})\phi _{b}^{(3-k)/b} & k=bn+3,\;n\in \mathbb{N} \\
0 & \mathrm{otherwise,}%
\end{array}%
\right.  \label{metallic_distrib}
\end{equation}%
where we can see explicitly which values of $k$ are not present for a given
value of $b$. This and other connectivity properties can be worked out from
the inflation process of the graphs. See Fig. \ref{metallicinflation}.

\bigskip We will center our attention on the first diagonal $d=1$. For every
$b$, the node positions on the first diagonal, $n=1,2,3,...$, are
\begin{equation}
N(n;1)=F_{2n},  \label{Nd}
\end{equation}%
that with the use of the generalized Binet formula
\[
F_{n}=\frac{1}{\sqrt{b^{2}+4}}\left[ \phi _{b}^{n}-\left( \frac{-1}{\phi _{b}%
}\right) ^{n}\right] \approx \frac{\phi _{b}^{n}}{\sqrt{b^{2}+4}},
\]%
can be written as

\begin{equation}
N(n;1)\approx \frac{1}{\sqrt{b^{2}+4}}\phi _{b}^{2n}=\frac{1}{\sqrt{b^{2}+4}}%
\phi _{b}^{2}\phi _{b}^{2n-2}=C_{b}\phi _{b}^{2n-4},  \label{N_phi}
\end{equation}%
where the position $n=1$ is

\begin{equation}
N(1;1)=F_{2} \approx \frac{1}{\sqrt{b^{2}+4}}\phi _{b}^{2} \equiv C_{b}.
\label{N(2)}
\end{equation}%
We note that the connectivity of the first node is $k(n=1)= b+3$ and in
general $k(n)=b+3+2b(n-1)$, $n\geq 2$. As before we redefine the
connectivities such that the degree is zero at the initial position $n=1$, $%
k(n)=2b(n-1)$, $n=1,2,3,...$ Following the same procedure as in Section 4,
from Eq. (\ref{N_phi}) we have

\begin{equation}
k(N(n;1))=2b(n-1)=\ln \left( \frac{N}{C_{b}}\right) ^{\frac{1}{\ln \phi _{b}}},
\label{degreeb}
\end{equation}%
and use of it in the sensitivity $\xi (N(n;1))\equiv \exp (k(n))$ yields

\begin{equation}
\xi (N(n;1))=\left( \frac{N}{C_{b}}\right) ^{\frac{1}{\ln \phi _{b}}}.
\label{sensitivityb}
\end{equation}%
Since all the features required for the $q$-deformation described in Section
4 are present for general $b$, we obtain for the generalized Lyapunov
exponent the expression
\begin{equation}
\lambda _{q}^{(b)}(1)=\frac{1}{N-C_{b}}\frac{\left( \frac{N}{C_{b}}\right) ^{%
\frac{1-q}{\ln \phi _{b}}}-1}{1-q} =\frac{1}{C_{b}\ln \phi _{b}},  \label{lambdaqb}
\end{equation}%
where $q=1-\ln \phi _{b}$. Likewise, the contents of Section 5 can also be reproduced for general $b$
with the result that%
\begin{equation}
h_{q}\left[ \pi (N)\right] =\lambda _{q}^{(b)}(1).  \label{pesinb}
\end{equation}

\section{Summary and discussion}

At the quasiperiodic onset of chaos the HV method leads to a self-similar
network with a structure illustrated by the related periodic networks
obtained from the sequence of attractors of finite periods along the route
to chaos \cite{luque6}. Under the HV algorithm many nearby trajectory
positions lead to the same network, since only when the values of trajectory
positions cross a threshold the corresponding node increases its degree with
new links. (See the succinct definition of the algorithm and the top panel in
Fig. \ref{lyapunovcirclemap-2}). Therefore trajectories off the attractor
but close to it transform into the same network structure. As we have seen
the fluctuations of the degree capture the anomalous but basic behavior of
the fluctuations of the sensitivity to initial conditions at the transition
to chaos \cite{robledo1}. The graph-theoretical analogue of the sensitivity was identified
as $\exp (k)$ while the amplitude of the variations of $k$ grows
logarithmically with the number of nodes $N$. These deterministic
fluctuations are described by a discrete spectrum of generalized
graph-theoretical Lyapunov exponents that are shown to relate to an
equivalent spectrum of generalized entropy growth rates, yielding a set of
Pesin-like identities. This behavior is similar to what was observed for the case of the more
straightforward period-doubling accumulation point \cite{luque5}. The
definitions of these quantities involve a scalar deformation of the ordinary
logarithmic function that ensures their linear growth with the number of
nodes. Therefore the entropy expression involved is extensive and of the
Tsallis type with a precisely fixed value of the deformation index $q$, $q=1-\ln \phi _{b}$,
where $\phi _{b}$ is the inverse of the irrational (dressed) winding number.

We have considered special families of time series and converted each into a
network, each family consists of the trajectories associated with an
attractor at the quasiperiodic transition to chaos of circle maps. The
attractors studied are defined by a winding number given by a quadratic
irrational or, equivalently, by a pure periodic continued fraction. Each
winding number singles out a specific route to chaos. Amongst these we
described in some detail the so-called golden route, but also we have shown
results for those known as the silver and bronze routes \cite{luque6}. See
Figs. \ref{farey} and \ref{expansion}. The HV algorithm proved to be capable
of generating a single network that contains the scaling and entropic
properties of the trajectories associated with each attractor. The results
presented here are of the same kind as those obtained for the
period-doubling route to chaos \cite{luque5} suggesting that the HV networks
associated with the onset of chaos are useful for describing the universal
properties at these special systems. The Pesin identity is a reflection of a
basic connection between BG statistical mechanics and chaos so that our
results provide elements for an analogous connection for the case of
nonergodic and nonmixing dynamics at vanishing ordinary Lyapunov exponent.

\textbf{Acknowledgements.} \ We acknowledge financial support by the
Comunidad de Madrid (Spain) through Project No. S2009ESP-1691 (B.L.),
support from CONACyT \& DGAPA (PAPIIT IN100311)-UNAM
(Mexican agencies) (A.R.).


\begin{thebibliography}{99}
\bibitem{strogatz1} S.H. Strogatz, Nonlinear Dynamics and Chaos: With
Applications to Physics, Biology, Chemistry, and Engineering, Perseus Books
Publishing, LLC, Reading, 1994.

\bibitem{dorfman1} J.R. Dorfman, An Introduction to Chaos in Nonequilibrium
Statistical Mechanics, Cambridge University Press, Cambridge, 1999.

\bibitem{luque1} L. Lacasa, B. Luque, F. Ballesteros, J. Luque, J.C. Nu\~{n}%
o, Proc. Natl. Acad. Sci. USA 105 (2008) 4973.

\bibitem{luque2} B. Luque, L. Lacasa, J. Luque, F. Ballesteros, Phys. Rev. E
80 (2009) 046103.

\bibitem{luque3} B. Luque, L. Lacasa, F. Ballesteros, A. Robledo, PLoS ONE 6
(9) (2011).

\bibitem{luque4} B. Luque, L. Lacasa, F. Ballesteros, A. Robledo, Chaos 22
(2012) 013109.

\bibitem{luque5} B. Luque, L. Lacasa, A. Robledo, Phys. Lett. A 376, 362
(2012).

\bibitem{luque6} B. Luque, A.M. N\'{u}\~{n}ez, F. Ballesteros, A. Robledo,
J. Nonlinear Sci. 23, 335 (2013).

\bibitem{luque7} A.M. N\'{u}\~{n}ez, B. Luque, L. Lacasa, J. P. G\'{o}mez,
A. Robledo, Phys. Rev. E 87, 052801 (2013).

\bibitem{hilborn1} R.C. Hilborn: Chaos and Nonlinear Dynamics. Oxford
University Press, New York (1994).

\bibitem{baldovin1} F. Baldovin, A. Robledo, Phys. Rev. E 69 (2004)
045202(R).

\bibitem{mayoral1} E. Mayoral, A. Robledo, Phys. Rev. E 72 (2005) 026209.

\bibitem{pesin1} Y.B. Pesin, Russian Math. Surveys 32 (1977) 114.

\bibitem{Crutchfield} J.P. Crutchfield, K. Young, Phys.Rev. Lett. 63, 105 (1989).

\bibitem{zhang06}
J. Zhang, M. Small, Phys. Rev. Lett. 96, 238701 (2006).

\bibitem{kyriakopoulos07} F. Kyriakopoulos and S. Thurner,
Lect. Notes in Comput. Sci. 4488, 625 (2007).

\bibitem{xu08} X. Xu, J. Zhang, and M. Small, Proc. Natl. Acad. Sci. USA, 105, 19601 (2008).

\bibitem{donner10}
R. V. Donner, Y. Zou, J. F. Donges, N. Marwan, and J. Kurths, New J. Phys. 12, 033025 (2010).

\bibitem{donner11}
R. V. Donner et al., Int. J. Bif. Chaos 21, 1019 (2010)

\bibitem{donner11-2}
R. V. Donner et al., Eur. Phys. J. B 84, 4, 653 (2011).

\bibitem{campanharo11}
A. S. L. O. Campanharo, M. I. Sirer, R. D. Malmgren, F. M. Ramos, L. A. N.Amaral,
PLoS ONE 6 (2011).

\bibitem{robledo1} H. Hern\'{a}ndez-Salda\~{n}a, A. Robledo, Physica A370,
286 (2006).

\bibitem{Bandt}
C. Bandt, B. Pompe, Phys. Rev. Lett. 88 (2002) 174102.

\bibitem{Landau} L.D. Landau, Dokl. Akad. Nauk SSSR 44, 339 (1944).

\bibitem{Ruelle} D. Ruelle and F. Takens, Commun. Math. Phys. 20, 167 (1971).

\bibitem{Shenker} S.J. Shenker, Physica D 5, 405 (1982).

\bibitem{Kadanoff} M.J. Feigenbaum, L.P. Kadanoff, and S.J. Shenker, Physica D
5, 370 (1982).

\bibitem{Rand} D. Rand, S. Ostlund, J. Sethna, and E.D. Siggia, Phys. Rev.
Lett. 49, 132 (1982).

\bibitem{Rand2} D. Rand, S. Ostlund, J. Sethna, and E.D. Siggia, Physica D 8,
303 (1983).

\bibitem{conway1} E.R. Berlekamp, J.H. Conway and R.K. Guy (1982), Winning
Ways (two volumes), Academic Press, London.

\bibitem{fraenkel1} A. S. Fraenkel, Theoretical Computer Science 282 (2002) 271–284.

\bibitem{Koelink} E. Koelink, W. van Assche, Proc. AMS 137, 5 (2009) 1663-1676.

\bibitem{robledo2} A. Robledo, Physica A 370, 449 (2006).

\bibitem{latora1} V. Latora, M. Baranger, Phys. Rev. Lett. 82 (1999) 520.

\end{thebibliography}
\end{document}